\documentstyle[natbib2, epsfig,natbibmnfix,epstopdf]{mn2e}

\def\msun{{\,{\rm M}_\odot}}
\def\kms{km$\,$s$^{-1}$}
\def\ergs{erg$\,$s$^{-1}$}
\def\simlt{\lower.5ex\hbox{$\; \buildrel < \over \sim \;$}}
\def\simgt{\lower.5ex\hbox{$\; \buildrel > \over \sim \;$}}

\title{Variable accretion and emission from the stellar
winds in the Galactic centre} 

\author[J.\ Cuadra et al]{Jorge Cuadra$^{1,2}$\thanks{e-mail: 
jcuadra@jilau1.colorado.edu}, 
Sergei Nayakshin$^{3}$, Fabrice Martins$^{4}$
\\
$^{1}$JILA, University of Colorado and National Institute of Standards and Technology, Boulder, CO 80309-0440, USA\\
$^{2}$Max-Planck-Institut f\"ur Astrophysik, D-85741 Garching, Germany\\ 
$^{3}$Department of Physics \& Astronomy, University of Leicester, LE1 7RH, UK\\ 
$^{4}$Max-Planck-Institut f\"ur extraterrestrische Physik, D-85741 Garching, Germany
}

\begin{document}

\date{Accepted XXX. Received XXX; in original form XXX}

\pagerange{\pageref{firstpage}--\pageref{lastpage}} \pubyear{2007}

\maketitle

\label{firstpage}

\begin{abstract}

We present numerical simulations of stellar wind dynamics in the
central parsec of the Galactic centre, studying in particular the
accretion of gas on to Sgr~A*, the super-massive black hole. Unlike
our previous work, here we use state-of-the-art observational data on
orbits and wind properties of individual wind-producing stars. Since
wind velocities were revised upwards and non-zero eccentricities were
considered, our new simulations show fewer clumps of cold gas and no
conspicuous disc-like structure.  The accretion rate is dominated by a
few close `slow wind stars' ($v_{\rm w} \le 750\,$\kms), and is
consistent with the Bondi estimate, but variable on time-scales of
tens to hundreds of years.  This variability is due to the stochastic
in-fall of cold clumps of gas, as in earlier simulations, and to the
eccentric orbits of stars. The present models fail to explain the high
luminosity of Sgr~A* a few hundred years ago implied by {\it Integral}
observations, but we argue that the accretion of a cold clump with a
small impact parameter could have caused it. Finally, we show the
possibility of constraining the total mass-loss rate of the `slow wind
stars' using near infra-red observations of gas in the central few
arcseconds.

\end{abstract}


\section{Introduction}

Sgr~A* is identified with the $M_{\rm BH} \sim 3.5 \times 10^6 \msun$
super-massive black hole (SMBH) at the centre of our Galaxy
\citep[e.g.,][]{Schoedel02, Ghez05}. By virtue of its proximity,
Sgr~A* may play a key role in the understanding of Active Galactic
Nuclei (AGN). Unlike any other SMBH, observations reveal in detail the origin of
the gas in the vicinity of Sgr~A*. This information is essential
for the accretion problem to be modelled self-consistently.

One of Sgr~A* puzzles is its very low luminosity with respect to
estimates of the accretion rate. Young massive stars in the inner
parsec of the Galaxy emit in total $\sim 10^{-3} \msun\,$yr$^{-1}$,
filling this region with hot gas. From {\em Chandra} observations, one
can measure the gas density and temperature around the inner
arcsecond\footnote{One arcsecond ($1''$) corresponds to $\sim
  0.04\,$pc, $\sim 10^{17}$cm, or $\sim 10^5 R_{\rm S}$ for Sgr~A*.}
and then estimate the Bondi accretion rate \citep{Baganoff03}. The
expected luminosity is orders of magnitude higher than the measured
$\sim 10^{36}\,$erg$\,$s$^{-1}$.

The hot gas, however, is continuously created in shocked winds
expelled by tens of young massive stars near Sgr~A*, and the stars
themselves appear to be distributed in two discs
\citep{Genzel03a, Paumard06}. The situation then is far more complex than in the
idealised -- spherically symmetric and steady state -- Bondi model. An
alternative approach is to model the gas dynamics of stellar winds,
assuming that the properties of the wind sources are known
\citep{Coker97, Rockefeller04, Quataert04,
  Moscibrodzka06}. These calculation however did not include the motion of the stars.

\cite{CNSD05, CNSD06} modelled the wind accretion onto Sgr~A*, for the
first time allowing the wind-producing stars to move, 
and showed the important influence of the orbits on the accretion. In
addition, they found that the winds create a complex two-phase medium
and that the accretion rate has a strong variablity on time-scales of
tens to hundreds of years. The stars, however, were modelled as a
group of sources whose features broadly reproduced the observed
distribution of orbits and mass loss properties. In this paper we
present our new simulations that treat the stellar population more
realistically. We now use the stellar positions and velocities as
determined by \cite{Paumard06}, and the wind properties derived by
\cite{Martins07} from the analysis of individual stellar spectra.

This paper starts with a description of our input parameters and a brief
account of the simulation method in \S~\ref{simul}. In \S~\ref{acc} we
characterise the simulated accretion flow in terms of origin, angular momentum, time-variability, and
expected X-ray emission. The morphology of the gas on a larger scale and its
expected atomic line emission are then explored in \S\S~\ref{morph} and
\ref{line}. We finally discuss our results in \S~\ref{discuss}.

\section{Simulations}
\label{simul}

The simulations were ran using the method described and tested in
detail by \cite{CNSD06}. We use the SPH/$N$-body code {\sc Gadget-2}
\citep{Springel05b} to simulate the dynamics of stars and gas in the
gravitational field of the SMBH. { To model the stellar orbits more accurately, we also
include the gravitational potential of the (old) stellar cusp, as
determined by  \cite{Genzel03a}. } 
The gas hydrodynamics
are solved with the SPH \citep[smoothed particle hydrodynamics;
e.g.,][]{Monaghan92} formulation, in which the gas is represented by a
finite number of particles that interact with their neighbours. { We
include optically thin radiative
cooling}. The SMBH is modelled as a `sink' particle
\citep{Bate95,Springel05a}, with all the gas passing within a given
distance from it ($0.05''$ in the present simulations) disappearing
from the computational domain. To model the stellar winds, new gas
particles are continously created around the stars.

As wind sources, we include 30 of the stars that \cite{Paumard06} identify as
Wolf--Rayet's (see Table~\ref{table:winds}). The remaining two stars, 3E and
7SE2, were not included in the models because of the poor constraints on their
orbits. None of these stars is very close to the black hole, nor {  is there any reason why they should possess larger mass loss rates compared to the other stars}\footnote{7SE2 is a WC9 star very similar to 7W which has $\dot{M} = 10^{-5} \msun\,$yr$^{-1}$. 3E is a WC5/6 star and \cite{Hillier99} derived
$\dot{M} = 1.5 \times\ 10^{-5}\msun\,$yr$^{-1}$ for a WC5 star.}. Consequently, we do not expect these two stars to have a strong influence on the accretion rate.

\subsection{Stellar wind data}

We used the wind properties derived for 18 of the mass-losing stars by
\cite{Martins07}. In that study, $H$ and $K$ band spectra of the
Wolf--Rayet stars in the central parsec of the Galaxy were analysed by
means of state-of-the-art atmosphere models. Mass loss rates ($\dot
M_{\rm w}$) and terminal wind velocities ($v_{\rm w}$) were derived
from the strength and width of emission lines. The models assumed
inhomogeneous (clumpy) winds, which lead to lower mass loss rates for the 8
stars previously analysed by \cite{Najarro97} by means of homogeneous models.
In addition, wind velocities of stars
displaying P-Cygni profiles were found to be larger than in
\cite{Paumard01} because the latter authors used only the emission
part of the P-Cygni profile to estimate the terminal velocity of the
winds.

For the 12 remaining stars, those not analysed in detail by
\cite{Martins07}, we set their wind parameters by simply using
those of similar stars that were properly studied. Table~\ref{table:winds}
shows the list of stars we use with their wind properties.

\begin{table}
\caption{Mass-losing stars and wind properties used in this paper.}
\begin{tabular} {r||l|r|r|r|}
\hline
ID&Name \ddag&$v_{\rm w}$&$\dot M_{\rm w}$&Note\\
&&\kms&$\msun\,$yr$^{-1}$&\\
\hline
19&16NW&     600&   1.12$\times 10^{-5}$&1\\
20&16C&     650&   2.24$\times 10^{-5}$&1\\
23&16SW&     600&   1.12$\times 10^{-5}$&2\\
31&29N&    1000&   1.13$\times 10^{-5}$&3\\
32&16SE1&    1000&   1.13$\times 10^{-5}$&3\\
35&29NE1&    1000&   1.13$\times 10^{-5}$&3\\
39&16NE&     650&   2.24$\times 10^{-5}$&4\\
40&16SE2&    2500&   7.08$\times 10^{-5}$&1\\
41&33E&     450&   1.58$\times 10^{-5}$&1\\
48&13E4&    2200&   5.01$\times 10^{-5}$&1\\
51&13E2&     750&   4.47$\times 10^{-5}$&1\\
56&34W&     650&   1.32$\times 10^{-5}$&1\\
59&7SE&    1000&   1.26$\times 10^{-5}$&1\\
60&--&     750&   5.01$\times 10^{-6}$&5\\
61&34NW&     750&   5.01$\times 10^{-6}$&1\\
65&9W&    1100&   4.47$\times 10^{-5}$&1\\
66&7SW&     900&   2.00$\times 10^{-5}$&1\\
68&7W&    1000&   1.00$\times 10^{-5}$&1\\
70&7E2&     900&   1.58$\times 10^{-5}$&1\\
71&--&    1000&   1.13$\times 10^{-5}$&3\\
72&--&    1000&   1.13$\times 10^{-5}$&3\\
74&AFNW&     800&   3.16$\times 10^{-5}$&1\\
76&9SW&    1000&   1.13$\times 10^{-5}$&3\\
78&B1&    1000&   1.13$\times 10^{-5}$&3\\
79&AF&     700&   1.78$\times 10^{-5}$&1\\
80&9SE&    1000&   1.13$\times 10^{-5}$&3\\
81&AFNWNW&    1800&   1.12$\times 10^{-4}$&1\\
82&Blum&    1000&   1.13$\times 10^{-5}$&3\\
83&15SW&     900&   1.58$\times 10^{-5}$&1\\
88&15NE&     800&   2.00$\times 10^{-5}$&1\\
\hline
\label{table:winds}
\end{tabular}
\\Notes:\\
(\ddag) IDs and names from \cite{Paumard06}.\\
(1) From \cite{Martins07}.\\
(2) Use 16NW.\\
(3) Use the average of 7W and 7SE.\\
(4) Use 16C.\\
(5) Use 34NW.\\
\end{table}

\subsection{Orbital data}

For each star we take the current 3D velocity and the position in the
sky determined by \cite{Paumard06}. The $z$-coordinate, i.e., its
distance from Sgr~A* projected along the line of sight, can be chosen
using different assumptions for the orbital distribution.  We tried a
range of reasonable assumptions, described in
\S~\ref{sec:orbits} below.

\subsubsection{Different orbital configurations}
\label{sec:orbits}

\paragraph{Almost circular orbits}
\label{sec:circ}

The simplest assumption one can make for the stellar orbits is to say
they are circular. However, for a given set of values $\vec v, x, y, M_{\rm BH}$
(3D velocity, 2D position, and central mass), it is not possible in
general to find a solution $z$ that satisfies at the same time the two
requirements for a circular Keplerian orbit: $v^2 = GM_{\rm BH}/r$ and
$\vec v \cdot \vec r = 0$, where $\vec r = (x,y,z)$. Instead, we look
for the value of $z$ that minimises the eccentricity, $e = \sqrt{1 +
  (2 \ell^2\epsilon)/(G^2M_{\rm BH}^2)}$, where $\vec \ell = \vec r
\times \vec v$ and $\epsilon = v^2/2 - GM_{\rm BH}/r$ are the angular
momentum and energy per unit mass of the orbiting star, respectively.
For the first run, {\sc min-ecc}, we set the current
$z$-coordinate to that value.

\paragraph{One stellar disc}
\label{sec:1disc}

\cite{Levin03} found that many of the young stars in the Galactic
centre, those rotating clockwise in the sky, have velocity vectors
that lie in a common plane. They interpreted this as a signature that
these stars are orbiting Sgr~A* in a disc. From updated observations,
\cite{Beloborodov06} estimated the thickness of this disc to be only
about $10^\circ$ and then calculated the most likely $z$-coordinate
for its stars. In our second orbital configuration, {\sc 1disc}, we used the
$z$-coordinate calculated in this way by \cite{Beloborodov06} for the
stars they identified as disc members, while for the rest of the stars
we use our previous assumption of low eccentricity orbits
(\S~\ref{sec:circ}).

\paragraph{Two stellar discs}
\label{sec:2discs}

\cite{Genzel03a} realised that the majority of the young stars in
the Galactic centre are actually confined to two almost perpendicular
discs. The second disc, with stars rotating counter-clockwise in the
sky, however, is not that well defined, being two times thicker than
the clockwise system \cite[see also \citealt{Lu06}]{Paumard06}. It is
not possible to obtain a robust estimate of the $z$-coordinate in this
case. For definitiveness we simply set $z = - (x n_x + y n_y) / n_z$
for this third orbital configuration, {\sc 2discs},
where $(n_x, n_y, n_z)$ is the vector perpendicular to the best
fitting counter-clockwise disc\footnote{With this setting, two of the
  stars (16NW, 29N) would have acquired orbits with pericentres $<
  0.1''$, comparable to the size of our inner boundary. To avoid
  numerical problems, we changed their velocities within the error
  bars, putting them in orbits that do not take them so close to the
  black hole.}. For the clockwise rotating stars, we used the
\cite{Beloborodov06} $z$-coordinates, as described in
\S~\ref{sec:1disc}.

\subsubsection{The `mini star cluster' IRS 13E}

Two of the wind emitting stars belong to IRS 13E, a group of stars
located 0.13 pc away from Sgr~A* in projection. The velocities of its
components are remarkably similar, so the group appears to be gravitationally 
bound. Since the SMBH
tidal force would quickly disrupt such a group, it is believed that it
harbours more mass than what is observed as massive stars, perhaps in
the form of an intermediate-mass black hole \citep{Maillard04,
  Schoedel05, Paumard06}. To take this into account when calculating
the $z$-coordinate of its components, we replaced their individually
measured velocities with the average group motion reported by
\cite{Paumard06}. Moreover, we include in the simulations a
$350\,\msun$ `dark matter' particle to keep the group bound.

\subsubsection{Setting the initial conditions}

Once the $z$-coordinates are set using any of the assumptions
described above (\S~\ref{sec:orbits}), we ran
$N$-body calculations to evolve the orbits back in time for 1100
yr. The final stellar positions and velocities from those runs were
used as initial conditions for the winds simulations.\footnote{We made
  an additional run starting 3000 yr ago. No significant differences
  were found, so we concentrate on simulations starting 1100 yr ago in
  this paper.}

\section{Accretion on to Sgr~A*}
\label{acc}

\begin{figure}
\centerline{\epsfig{file=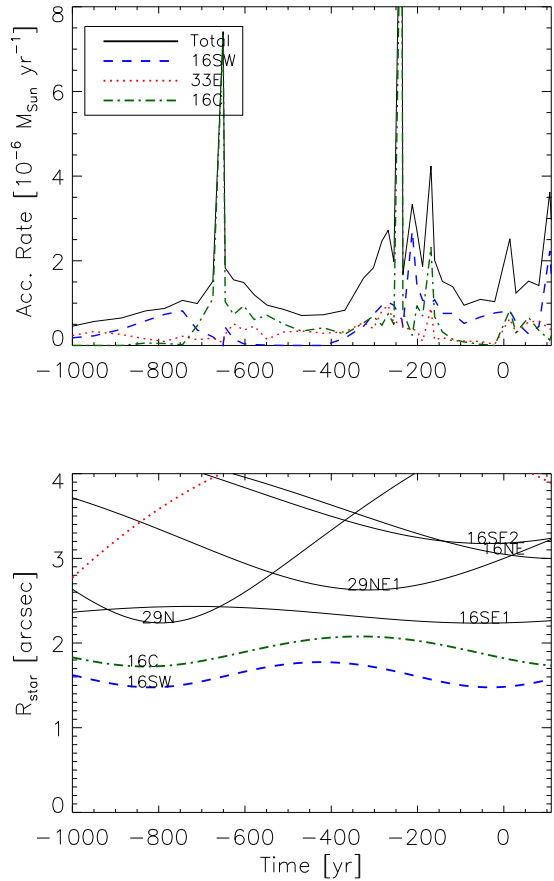,width=.49\textwidth}}
\caption{Top panel: Accretion rate as a function of time for run {\sc
min-ecc}. The accretion curve is created by sampling the accreted mass every
$\sim 30\,$yr and a time value of zero corresponds to the present time. The
black solid line shows the total accretion rate while the coloured broken
lines show the contribution of the three stars that dominated accretion on to
Sgr~A* during the simulation. Bottom panel: Distance from the innermost
stars to the black hole as a function of time for the same simulation. The
curves are labelled with the star names and the three most important stars are
shown with the same line properties as in the top panel.}
\label{fig:acc_circ}
\end{figure}

\begin{figure}
\centerline{\epsfig{file=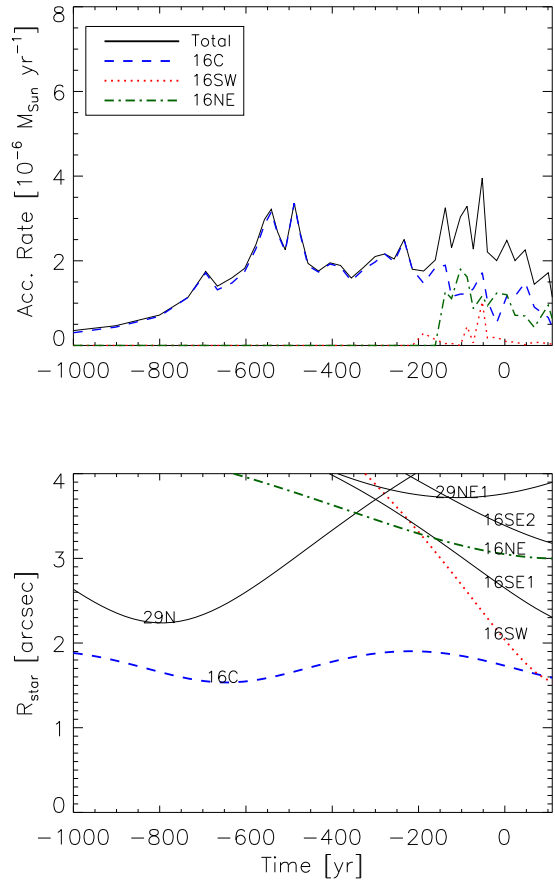,width=.49\textwidth}}
\caption{Same as Fig.~\ref{fig:acc_circ} for run {\sc 1disc}.}
\label{fig:acc_1disc}
\end{figure}

\begin{figure}
\centerline{\epsfig{file=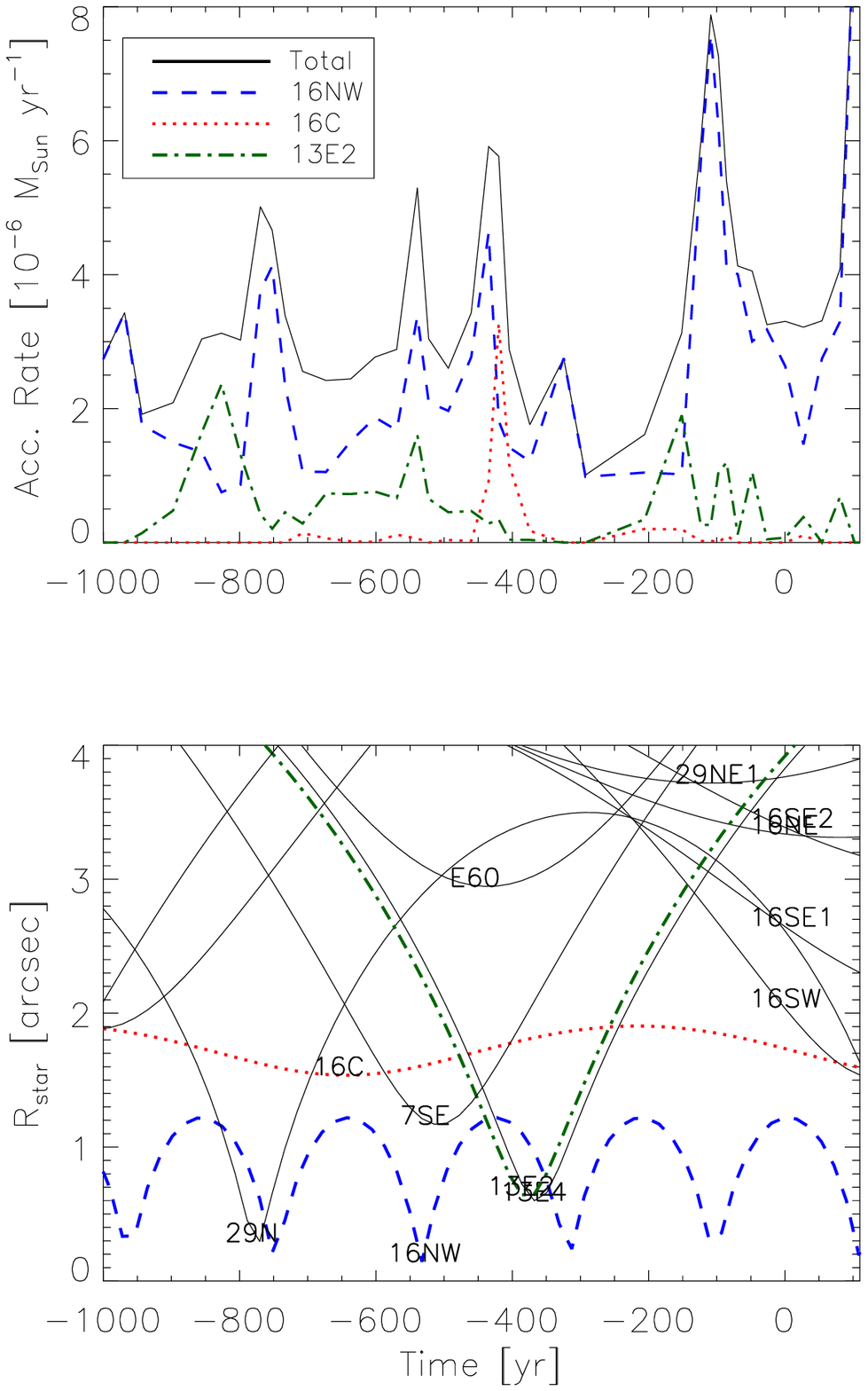,width=.49\textwidth}}
\caption{Same as Fig.~\ref{fig:acc_circ} for run {\sc 2discs}.}
\label{fig:acc_2discs}
\end{figure}

\subsection{Variability and origin of the accretion}

We define the accretion rate on to Sgr~A* as the rate at which mass enters
the inner boundary ($0.05''$) of our computational domain. The upper panel of
Fig.~\ref{fig:acc_circ} shows the so-defined accretion rate for run {\sc min-ecc}, where the orbits were made as circular as
possible. The lower panel
shows the distance from the innermost stars to the black hole as a function of
time for the same simulation. In this case most of the accreted material comes from the innermost
stars 16SW and 16C, whose orbits oscillate in the range $\sim1.5$--$2"$. These
stars have relatively slow wind velocities, $v_{\rm w} \approx
600\,$\kms. Additionally, star 33E, that was further away but has even slower
winds with $v_{\rm w}=450\,$\kms, makes an important contribution to the
accretion rate. The slow winds from these three stars are captured by
Sgr~A$^*$ more easily than winds from stars like 29N and 16SE1, which
were at comparable distances from the black hole, but have faster winds that
were not accreted at all.

Figures~\ref{fig:acc_1disc} and \ref{fig:acc_2discs} show the accretion rate
corresponding to runs {\sc 1disc} and {\sc 2discs}, respectively. The
accretion is again dominated by stars with wind velocities of at most
$750\,$\kms. The accretion rate history is however different between the
simulations. When the orbits are almost circular (Fig.~\ref{fig:acc_circ}), the accretion rate is more
stable, with only a few narrow peaks produced by {
the dense clumps resulting from the cooling of slow  winds (see \S~\ref{morph})}. 
When the stars are set in one or two discs, the orbits typically have
higher eccentricities (notice the variation in the distance vs.\ time plots),
changing the quantity of gas that can be captured by the black hole as a
function of time. This is especially true for run {\sc 2discs}, where the innermost star 16NW has quite an eccentric orbit. {  As a result, when this star approaches the pericentre of its orbit, a large fraction of its winds is captured directly by Sgr~A* 
within our numerical framework.}

\cite{CNSD06} used only circular orbits and found that, while the
total amount of accretion was dominated by hot gas (temperature $T > 10^7\,$K),
the variable accretion rate was mainly caused by the infall of cold
clumps. In these new simulations, the accretion is again dominated by
hot gas, but there is almost no accretion of cold gas at all. The reason for
this difference is most likely the eccentricity of the orbits -- closer to the
black hole the stars acquire high orbital velocities that increase the
total velocity of the wind, giving it a large kinetic energy which is
then thermalized. Only in the run with orbits closer to circular we
see a few sharp peaks in the accretion rate, originated by cold clumps
that survived the hot inner region and reached the black hole.
On the other hand, the accretion of hot gas shows larger variability than
in our previous calculations. 

The value of the accretion rate is of the order of a few
$\times10^{-6}\msun\,$yr$^{-1}$, consistent with the expectations from
the Bondi model \citep{Baganoff03}. This means that the reason for
Sgr~A* low luminosity lies in the physics of the inner accretion flow
that we cannot resolve with our simulations, and not in how much material is
captured by the black hole at distances $\sim 10^4 R_{\rm S}$.

\subsection{Accretion luminosity}\label{sec:xray}

We shall try to estimate the X-ray luminosity produced by the accretion flow
based on the accretion rate measured at the inner boundary $R= 0.05''$ of the
computational domain. Physically, gas that reached this point still needs a time 
of the order of the flow's viscous time, $t_{\rm visc}$, to reach Sgr A*. This 
effect smoothes out any variability in the accretion rate that proceeds
on time-scales $\Delta t < t_{\rm visc}$. Therefore, to calculate the
luminosity of the flow we first average the instantaneous accretion rate over
time intervals $t_{\rm visc}$.

From the standard accretion theory \citep{Shakura73}, the viscous
time-scale can be estimated as 
\begin{equation}
t_{\rm visc} = \frac{1}{ \alpha \Omega_{\rm K}} (\frac{R}{H})^2 =
6.8\,\hbox{yr} \, R_{0.05''}^{3/2} \alpha_{0.1}^{-1}  (\frac{R}{H})^2 ,
\end{equation}
{ where $\alpha = 0.1\alpha_{0.1}$ is the standard viscosity parameter, $\Omega_{\rm K}$ is the Keplerian orbital frequency, $R = 0.05" R_{0.05''}$ is the distance from the black hole, and $H$ is the disc thickness.}
If we take $\alpha=0.1$, and a geometrically thick disc, $H/R \approx
1$, appropriate for radiatively inefficient accretion, the viscous timescale can be as short as
$t_{\rm visc} \approx 5\,$yr. While for a thin disc this timescale can
be of course much longer, we concentrate in the $t_{\rm visc}
\sim 5\,$yr regime.

The 
radiative efficiency of the flow is highly uncertain. We therefore use two prescriptions for the X-ray luminosity of the accretion flow $L_{\rm X}$ as a function of the accretion rate $\dot M$,  
\begin{equation}
L_{\rm X} = 0.01 (\frac{\dot M}{\dot M + \dot
M_{\rm crit}})^\beta \dot M c^2, 
\end{equation}
where the parameter $\beta$ is set to either 1 or 2 and $c$ is the speed of light. 
{  This formula provides a smooth transition between the high accretion rate regime, where $L_{\rm X}$ is expected to be proportional to $\dot M$, and the low accretion rate limit, where the luminosity dependence on $\dot M$ is steeper \citep[see, e.g.,][about these two regimes]{Merloni03}}
The critical value of the accretion rate,
where it would switch {  between these regimes}, is set at $\dot M_{\rm crit} =
0.01 \dot M_{\rm Edd} \approx 7.7\times10^{-2}\msun\,$yr$^{-1}$, as suggested
by the study of different modes of accretion in AGNs
\citep[e.g.,][]{Maccarone03}.

\begin{figure}
\centerline{\epsfig{file=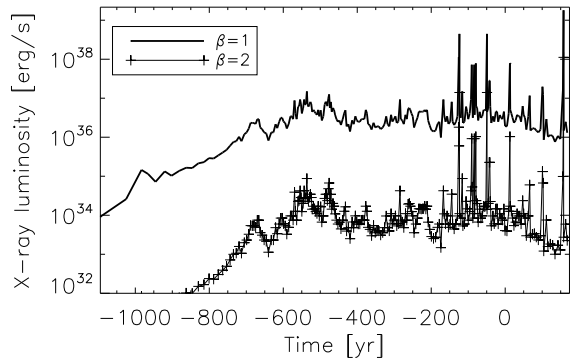, width=.48\textwidth}}
\caption{Luminosity from the accretion flow in run {\sc 1disc}. The solid and
crossed lines correspond to the estimates with $\beta=1,2$, respectively.}
\label{fig:acclum_belo}
\end{figure}

As an example, we show in Fig.~\ref{fig:acclum_belo} the resulting X-ray luminosity of run {\sc 1disc}. 
As can be seen from the figure, the choice $\beta = 1$
gives a typical luminosity for Sgr~A* that is too large compared with
the $\sim 2\times10^{33}\,$\ergs currently observed. The case $\beta = 2$
gives more reasonable results, with typical values $L_{\rm X} \sim
10^{33}$--$10^{34}\,$\ergs, { and is actually closer to the relation ($ L_{\rm X} \propto \dot m^{3.4}$) calculated for very inefficient accretion by \cite{Merloni03}.}

Interestingly, our simulations show that in the recent past Sgr~A* X-ray luminosity could have
reached more than $10^{37}\,$\ergs. On the observational side, \cite{Revnivtsev04} detected hard X-rays from molecular clouds in the vicinity of Sgr~A* with {\it Integral}, and interpreted them  
as reflection from a bright source, implying a
past luminosity of $\sim 10^{39}$\ergs\ for Sgr~A* that lasted at least a dozen
years. It seems difficult to reproduce such high values with our models,
especially because tweaking the viscous time-scale to make the peaks
higher would at the same time make them narrower than the required duration. 
{  However, low mass gas clumps are obviously under-resolved in our simulations at some level. It is possible that the in-fall of a particularly low angular momentum clump much closer in to Sgr~A* than our inner boundary radius would result in a flare with the observed characteristics (see \S~\ref{discuss}).}

\subsection{Angular momentum and circularisation radius}

\begin{figure}
\centerline{\epsfig{file=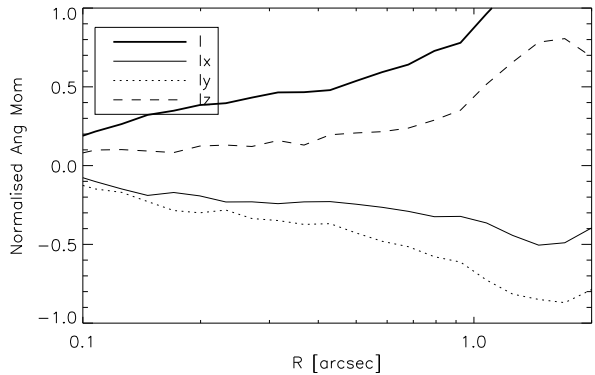, width=.48\textwidth}}
\caption{Spherically averaged angular momentum profile of the inner
  region of the accretion flow of the simulation {\sc min-ecc}. The profile
  is built using several snapshots in the range $t \approx 0 \pm
  37\,$yr.}
\label{fig:angmom_circ}
  \end{figure}

\begin{figure}
\centerline{\epsfig{file=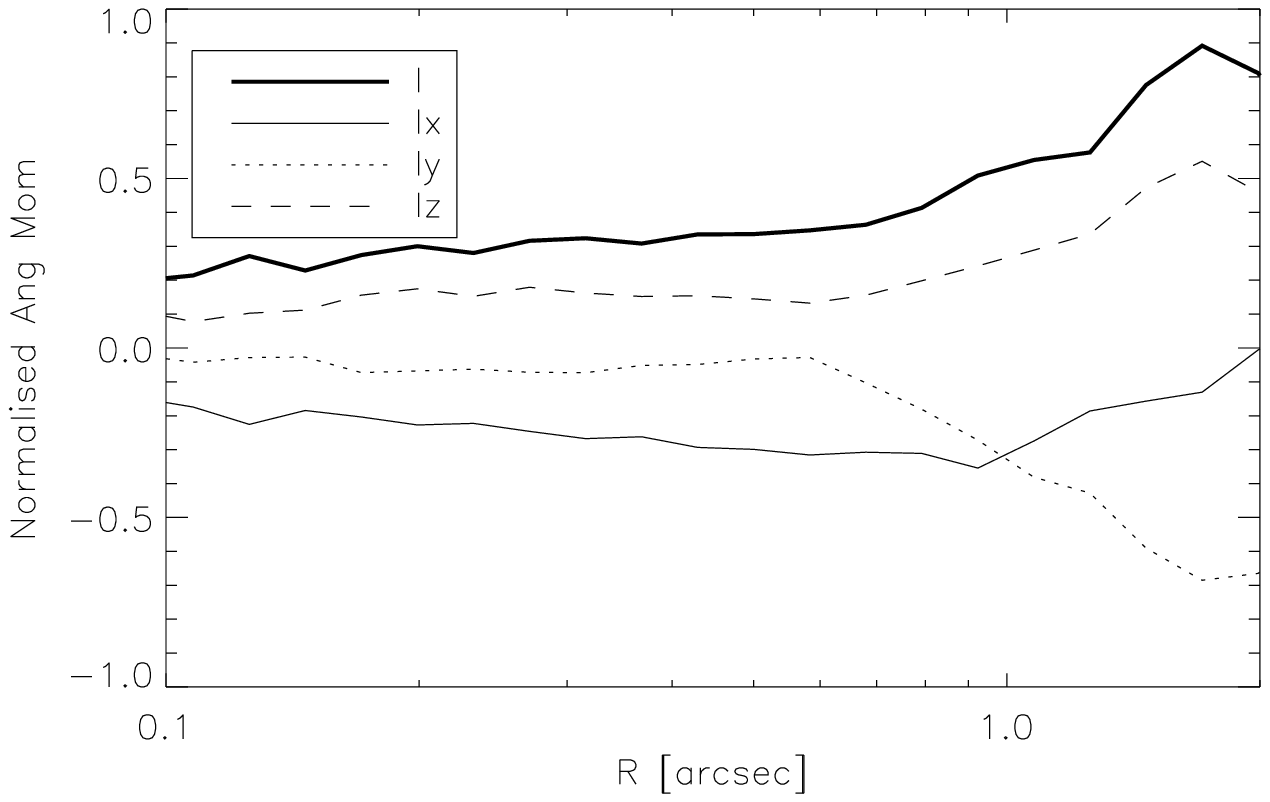, width=.48\textwidth}}
\caption{As Fig.~\ref{fig:angmom_circ}, but for run {\sc 1disc}.}
\label{fig:angmom_belo}
\end{figure}

\begin{figure}
\centerline{\epsfig{file=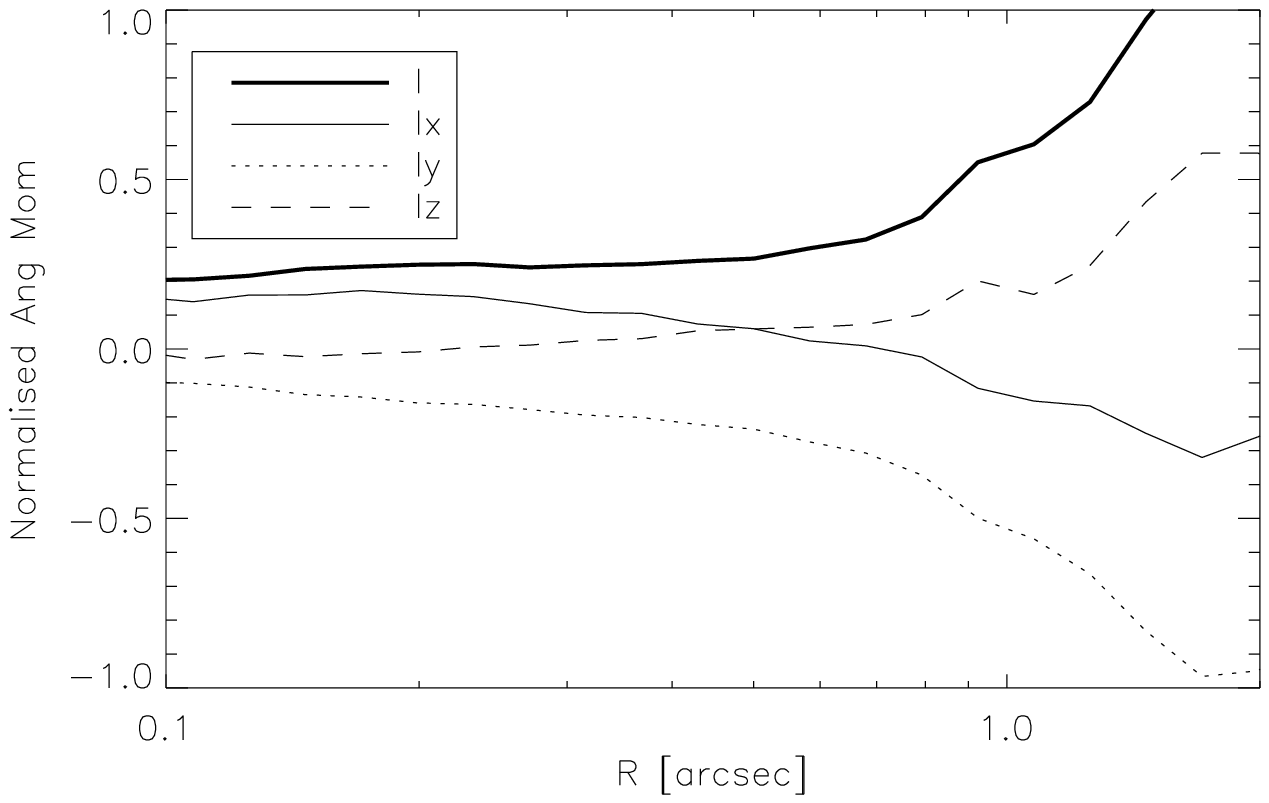, width=.48\textwidth}}
\caption{As Fig.~\ref{fig:angmom_circ}, but for run {\sc 2discs}.}
\label{fig:angmom_2discs}
\end{figure}

\begin{figure}
\centerline{\epsfig{file=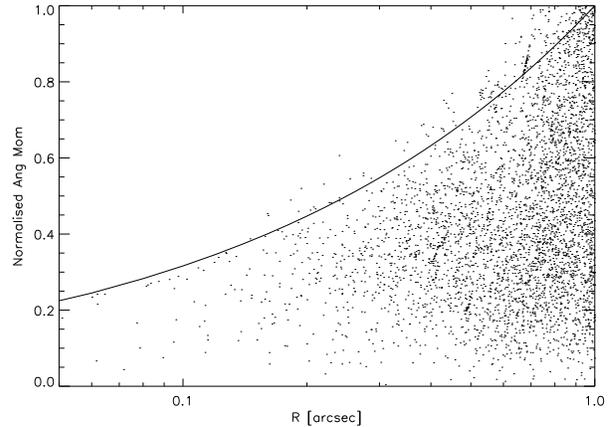,width=.48\textwidth}}
\caption{  Scatter plot showing the specific angular momentum of every SPH particle in the inner $1"$ for simulation {\sc 1disc} at present time ($t=0$). { Very few particles have angular momentum larger than the Keplerian value, shown with the solid line.}}
\label{fig:angmomdist}
\end{figure}

One important piece of information to characterise the accretion on to Sgr~A*
is the angular momentum of the gas that it accretes.
Figures~\ref{fig:angmom_circ}--\ref{fig:angmom_2discs} show the angular momentum profile of the three different simulation at $t \approx 0$, 
corresponding to the present time. The
gas close to our inner boundary has an average angular momentum
$\ell \approx 0.25$, in units where a circular Keplerian orbit at $R=1''$
would have $\ell = 1$. 
{  We use this value of the angular momentum to roughly estimate at which distance from the black hole the flow will circularise as $R_{\rm circ} = \ell^2$,
  which gives $R_{\rm circ} \sim 0.05'' \sim 5000 R_{\rm S}$ for our simulations. The exact value of circularisation radius is important for modelling the inner accretion flow, and we thus need to examine robustness of our results. 

 Figure~\ref{fig:angmomdist} shows the angular momentum values of all SPH particles in the inner $1"$ of the simulation domain. Not surprisingly, at every radius there are very few gas particles that exceed the local Keplerian angular momentum, shown with the solid curve. However, there is significant scatter, i.e., the angular momentum distribution of SPH particles has a significant spread about the mean time-averaged value plotted in Figures~\ref{fig:angmom_circ}--\ref{fig:angmom_2discs}. Without performing a much higher resolution simulation it is difficult to say whether this scatter is real or numerical. 
In the former case, this would imply that there is a range of angular momentum values for the gas entering the capture radius, and that $\ell \approx 0.25$ is indeed only the average value. 
We would expect then the gas to circularise over a range of radii, with the circularisation radius $R_{\rm circ} \sim 0.05''$ representing only a rough geometrical mean value.  A further complication is the absence of magnetic fields in our simulations, which could provide means for angular momentum transfer \citep[e.g.,][]{Balbus98, Proga03b}. Therefore, we conclude that our results on the circularisation radius value are only preliminary, and more work is needed to pin down its value and dependency on the stellar orbits.}

%\begin{figure}
%\centerline{\epsfig{file=velrad.eps, width=.48\textwidth}}
%\caption{Average radial velocity of the gas for the three different simulations.  The gas in average flows in once it reaches the inner arcsec.  Out of this region most of the gas outflows.}
%\label{fig:radvel}
%\end{figure}

We also plotted the different components of the gas angular momentum (Figs.~\ref{fig:angmom_circ}--\ref{fig:angmom_2discs}). While
the magnitude of the specific angular momentum of gas in all the three
simulations is similar near the inner boundary, its direction varies
significantly. These simulations thus predict different orientations for the
midplane of the accretion flow, which is not surprising given a rather
significant difference in stellar orbits between the simulations.

\begin{figure}
\centerline{\epsfig{file=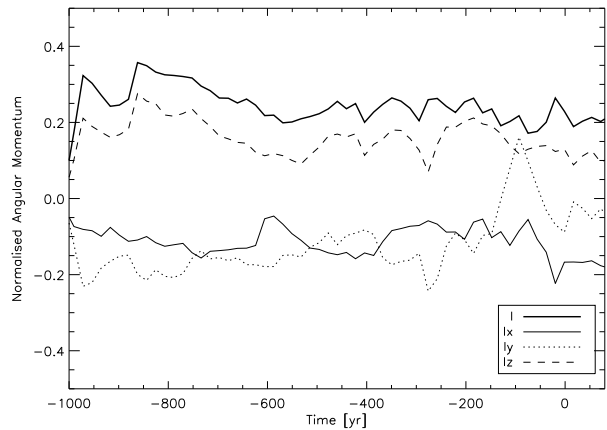, width=.48\textwidth}}
\caption{Average angular momentum of the inner $0.3"$ as a function of time from simulation {\sc 1disc}.}
\label{fig:angmomtime}
\end{figure}

The angular momentum also varies as a function of time, as the
geometry of the stellar system changes. Fig.~\ref{fig:angmomtime}
shows the average angular momentum of the gas in the inner $0.3"$ as a
function of time for the simulation {\sc 1disc}. Both the magnitude
and the orientation of the angular momentum change by up to a factor 2
on time-scales of tens of years. A sudden change in the angular
momentum vector, as that seen in this simulation at $t \approx
-100\,$yr, can strongly perturb the inner accretion flow and produce
an episode of enhanced accretion.

\section{Gas morphology}
\label{morph}

\begin{figure*}
\begin{minipage}[b]{.49\textwidth}
\centerline{\includegraphics[height=10cm]{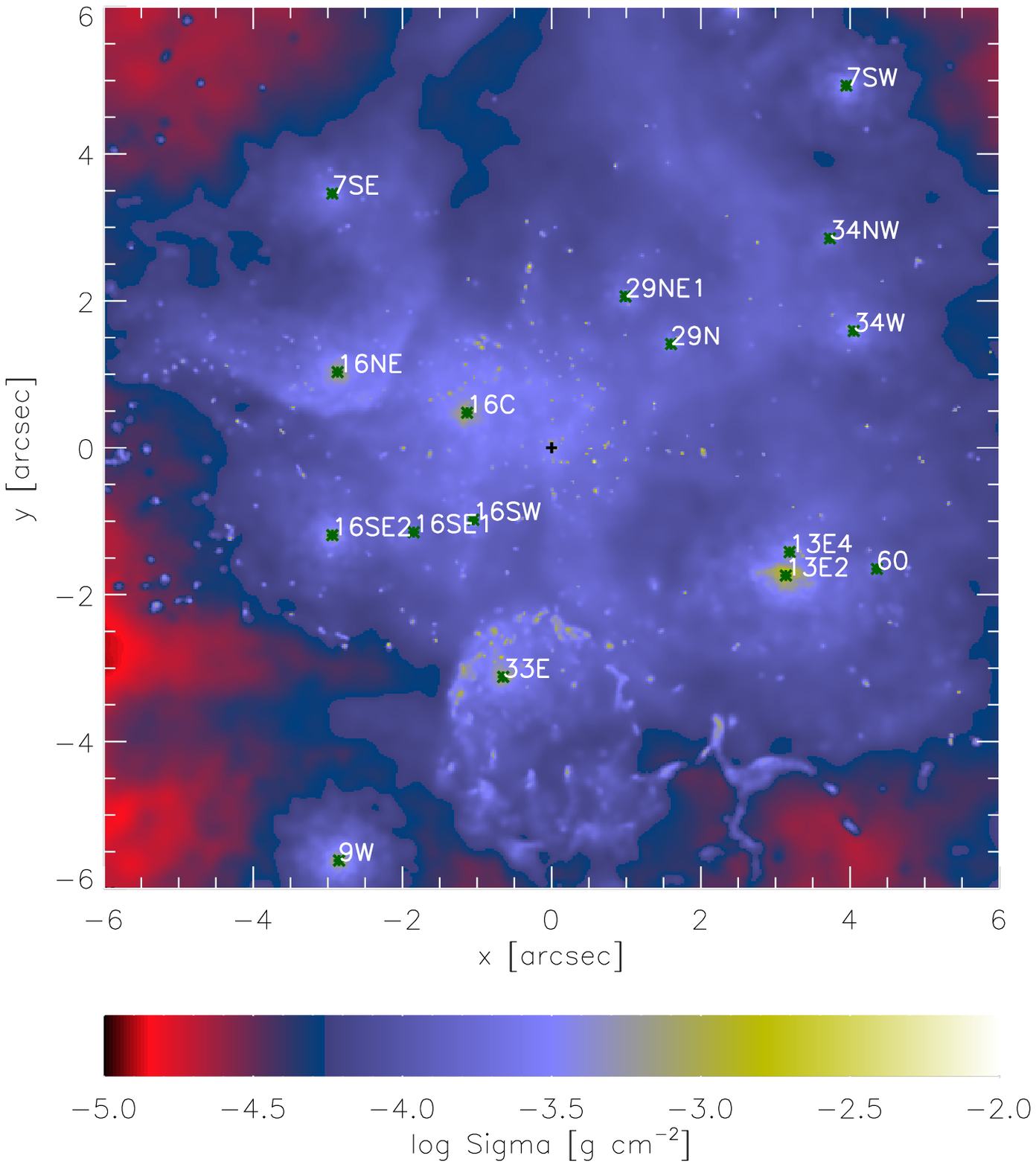}}
\end{minipage}
\begin{minipage}[b]{.49\textwidth}
\centerline{\includegraphics[height=10cm]{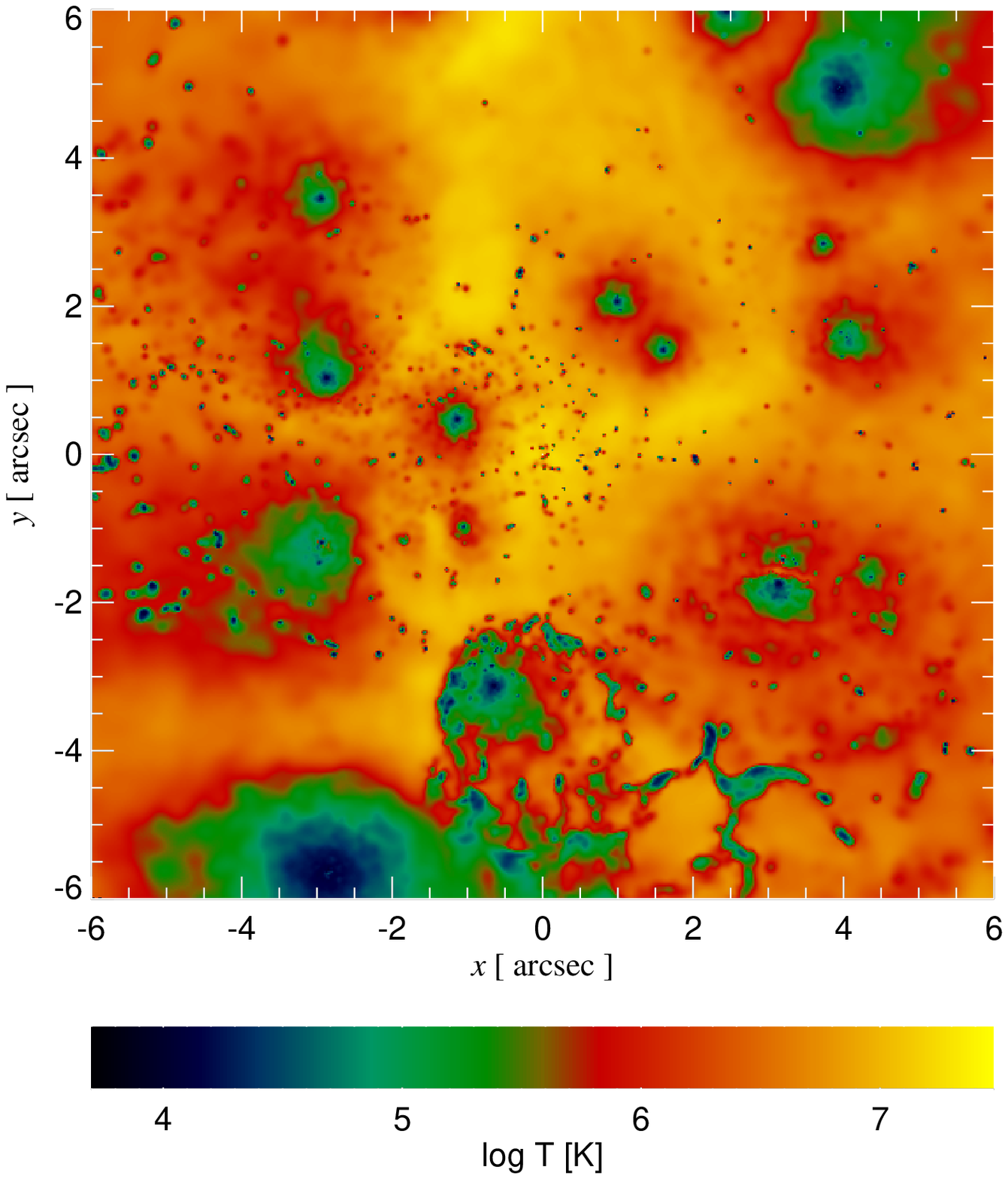}}
\end{minipage}
\caption{Gas morphology at the present time ($t=0$) from the simulation {\sc
  1disc}. Left panel: Column density of gas in the inner $6''$ of the
  computational domain, as it would be observed from Earth. Stars are shown
  with green symbols, with labels indicating their names.  Right panel:
  Averaged temperature of the same region. Notice the dense cold clumps
  forming around the slow-wind-emitting star 33E. Clumps also form in the
  region around the 13E group, where the slow winds from 13E2 collide with the
  faster ones comming from its neighbour 13E4. On the other hand, winds from
  the powerful WR star 9W have not collided yet with any other winds and
  remain cold but diffuse.}
\label{fig:largeview_belo}
\end{figure*}

Figure \ref{fig:largeview_belo} shows the resulting morphology of the
gas at present time from run {\sc 1disc}. { The other two simulations show no important differences, so we concentrate on this intermediate case.}
The cool and dense clumps originate from
the slow winds. When shocked, these slow winds attain a temperature of
only around $3\times10^6$ K, and, given the high pressure environment
of the inner parsec of the GC, cool radiatively over a time-scale
comparable to the dynamical time \citep{CNSD05}. On the other hand,
the fast winds do not produce much structure by themselves. This gas reaches
temperatures $> 10^7\,$K after shocking, and does not cool fast enough
to form clumps. This temperature is comparable to that producing X-ray
emission detected by {\em Chandra}.  Gas cooler than $\sim 10^7\,$K would be
invisible in X-rays due to the finite energy window of {\em Chandra}
and the huge obscuration across the Galactic plane.

The new simulations confirm our previous results that the winds create
a two-phase medium in the Galactic centre \citep{CNSD06}. The quantity
of cold gas, however, is very much reduced in the new simulations. We
also notice that there is no disc-like structure like the
one previously found \citep[Fig.~9 in][]{CNSD06}.

There are several reasons for the differences. The wind velocity of
the slow wind stars -- from whose winds the cold clumps are mostly
formed -- were revised upward from $300\,$\kms to typically
$650\,$\kms. As the radiative cooling strength for this gas is a
strong function of its initial velocity \citep{CNSD05}, 
a much smaller fraction of gas can cool and form
clumps. Additionally, the mass loss rates we use for the slow wind
stars were revised downward by a factor of a few compared to our previous
calculations, so there is even less material that could potentially
form clumps. 

\cite{An05} found that Sgr~A* flux at $\sim 1\,$GHz increased by a
factor $\sim 2$ from 1975 to 2003.{ These authors attribute the change
to a decrease in the free--free opacity produced by a
factor 9 change in the column-integrated density squared, $\int n^2 dl$. As it is clear from the density map in 
Fig.~\ref{fig:largeview_belo}, the motion of clumps of gas can easily
account for such a change in the obscuration.} Further observations
should be able to estimate the time-scales on which the change of
obscuration happens and therefore confirm the identification of cold
clumps with the obscuring material.

\section{Line emission}
\label{line}

To compare better the outcome of our simulations with actual observations, we
create emission maps of { strong near-infrared} atomic lines that are expected from gas at $T \sim
10^4\,$K.  In the simulations the minimum temperature is set to $T =
10^4\,$K. This is justified since the powerful UV radiation from the stars in
the region ensures that most of the gas remains ionised\footnote{For one
single star with ionising radiation rate $Q = 10^{48} Q_{48}\,$s$^{-1}$, the
Str\"omgren radius is $ 1.38\,$pc$\,Q_{48}^{1/3} n_2^{-2/3}$.}.

Only a limited comparison with the data is possible at this stage. Our
simulations do not include the Mini-spiral, a rather massive ($\sim 50 \msun$)
and large scale ionised gas feature{ composed of several dynamically
independent structures \citep[e.g.,][]{Paumard04}. While the
origin of the Mini-spiral gas is not quite clear, it seems to follow eccentric orbits
originating outside the inner parsec.} Fortunately for our study, there does not
seem to be much of the Mini-spiral material within few arc-seconds of Sgr~A*,
so we choose to concentrate on the inner region of the computational domain.
Our aim is to find out whether the gas that is produced by the stellar winds
would produce a level of emission that is too high compared to the observed
value. If that is the case, either the physics of our model is wrong, or the
input parameters we are using (stellar properties and orbits) should be
revised.

\begin{figure}
\centerline{\epsfig{file=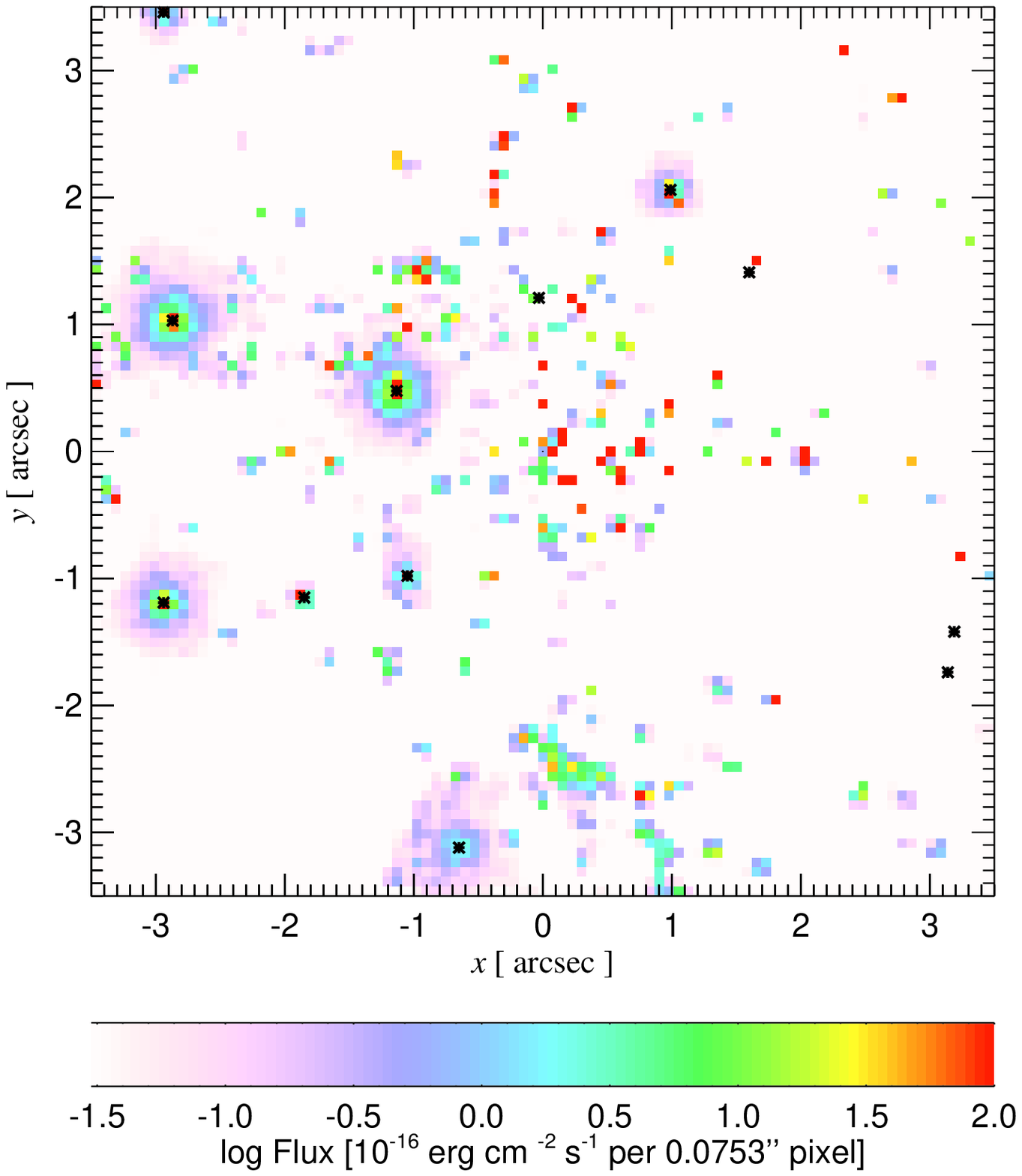,width=.49\textwidth}}
\caption{Pa$\alpha$ surface brightness from the stellar winds in run {\sc 1disc}, as it would be observed in the sky at present time. The units are those used by \cite{Scoville03}. Stars are shown as black asterisks. The level of emission is too low to be detected on top of the other gas complexes in the GC region.}
\label{fig:paa_belo}
\end{figure}

In particular, we create maps of Pa$\alpha$ emission and compare them with
the observations of \cite{Scoville03}. We calculate the luminosity per unit
volume as
\begin{equation}
 4 \pi j_{\rm Pa\alpha} = 6.41 \times 10^{-18}\,
{\rm erg\,cm}^{-3}\,{\rm s}^{-1} (\frac{T}{6000\,{\rm K}})^{-0.87} (\frac{n}{10^4\,{\rm cm}^{-3}})^2
\end{equation}
\citep{Osterbrock89} and
then integrate it along the line of sight. An example from simulation {\sc 1disc} is shown
in Fig.~\ref{fig:paa_belo}. Only a few pixels have a surface brightness
comparable to that measured in the Galactic centre inner few arc-seconds
\citep[their Fig.~7]{Scoville03}, making the present simulations compatible
with the observations. This is in contrast to our earlier simulations
\citep{CN06}, the analysis of which shows too much atomic line emission in the
near infrared. The difference is not unexpected as the earlier simulations produced much more
cold gas in the inner region (see \S \ref{morph}).

{ While a more detailed analysis is left to future work, it is clear
that in general the gas luminosity depends on the wind
properties. Once more robust orbital data is obtained, it may be
possible to constrain the properties of the stellar population, in
particular the total mass loss rate from `slow wind stars', using this
method that is complementary to the stellar spectra analysis
\citep[e.g.,][]{Martins07}.}

\section{Discussion}
\label{discuss}

We presented our new simulations of stellar wind dynamics in the Galactic
centre. We use state-of-the-art data on the stellar orbits
\citep{Paumard06} and stellar winds properties
\citep{Martins07}. Unfortunately, this does not eliminate the uncertainty in
the models completely. The $z$-coordinate of the stellar wind sources, i.e.,
the distance along the line of sight to the GC, cannot be obtained
observationally.  We therefore made several simulations each with a different
assumption about the dynamics of these stars, from almost circular
orbits to orbits confined in two planes.

The main result of our simulations, compared with results of \cite{CNSD05,
CNSD06}, is the much smaller quantity of cool gas at $T \simlt 10^4$ K. We
find no cool and geometrically thin disc formed from radiatively cooled
stellar winds. This is due to two factors. Firstly, the updated stellar wind
velocities are significantly higher, implying that radiative cooling time
becomes too long for the shocked winds to form cool clumps. Secondly, the
orbits of the important mass losing stars are less disc-like, { i.e, circular and coplanar, } compared with
most of our previous tests. 

On the other hand, similar to the results of \cite{CNSD05, CNSD06}, we find a
substantial time variability in the accretion rate histories for Sgr~A* in the
three different models explored here.  In the simulation with small
eccentricities ({\sc min-ecc}) the accretion is relatively constant, except
for the episodic in-fall of cold clumps that produce sharp peaks in the
accretion rate. When the orbits are preferentially in one or two planes
(simulations {\sc 1disc} and {\sc 2discs}), stars acquire higher
eccentricities. The accretion rate history is then strongly variable on a time-scale of tens to hundreds of years mainly due to stars on eccentric
orbits. The accretion rate peaks are nearly coincident with times when these
stars are at pericentres of their orbits. {  One should notice however that in the simulation 
{\sc 2discs} the star 16NW gets uncomfortably close to the inner boundary of the computational domain (Fig.~\ref{fig:acc_2discs}), so higher resolution studies are needed to quantitatively confirm the obtained accretion rate.}  {   Previous studies have also found substantial variability of the accretion flow, e.g.,  \cite{Proga03b}, using MHD models for the region closer to the central black hole \citep[see also][]{Moscibrodzka07}.
 }

Even though the accretion history looks quite different between the
simulations, the average accretion rates are the same within a factor of 2--3. In particular,
time-averaged accretion rates of all of the simulations are consistent
with the 
{  Bondi estimate. 
This is not unexpected. The capture radius of the wind is around $0.5"$ to $1"$. At these radii, the 
angular momentum of the gas is well below its Keplerian circular value. Therefore, rotational support 
of the flow is unimportant, and the Bondi approximation is valid at these radii. On the other hand, 
had we been able to resolve and properly model much smaller scales, the
angular momentum effects would be considerably more important. This is exactly the direction in which
this work needs to be extended to connect it with the radiatively-inefficient accretion flow 
(RIAF) models that have been developed for Sgr~A*. 
In these models only a small fraction of the energy of the accretion flow is radiated away, allowing them 
to successfully account for Sgr~A* dimness  \citep[e.g.,][]{Narayan02, Yuan03}. Also, most of the gas captured
by Sgr~A* on arc-second scales does not get accreted but rather outflows. In addition to improved
numerical resolution, magnetic fields need to be included to account for angular momentum transfer in the flow.}

The angular momentum of the accretion flow in the sub arc-second region
appears to be similar for the different orbital configurations we tried, 
with the average circularisation radius being of the order of $\sim
5\times 10^3$ Schwarschild radii. 
 {  This radius is similar to the size of the inner boundary of our simulations.  Moreover, its value is calculated from the angular momentum of the gas in the inner region of the computational domain, where the quantity of SPH particles may not be enough to yield robust results.  More work is then needed to pin down the exact value of the circularisation radius. 
 Higher numerical resolution -- lower SPH particle mass and a smaller value for the inner boundary radius -- may result in a smaller value for the circularisation radius. On the other hand, inclusion of magnetic fields, absent in our present simulations, will allow angular momentum to be transfered in the flow via the magnetorotational instability \citep[e.g.,][]{Balbus91, Balbus98}, perhaps allowing material with a higher value of angular momentum to be `accreted' within the inner boundary. Hence the estimate for the circularisation radius can in principle shift somewhat in either direction. }

Although the angular momentum magnitude is roughly the same, the orientation of the
accretion flow mid-plane is different in the three
simulations, since it is mostly determined by the orbits of a few of the innermost stars. Better measured velocities and constraints on the
accelerations \citep[e.g.,][]{Lu06} should be obtained during the next
years of observations, allowing us to improve the determination of the
3-dimensional orbits and get more definitive answers. Once 
the orbits are better constrained, one
could embark on a more ambitious higher resolution study of the inner
accretion flow in an attempt to connect it with the non-radiative
models mentioned above, and the 
observational constraints on the inner accretion flow orientation
 \citep[e.g.,][]{Meyer06,Trippe07}.

In this paper we already attempted some modelling of the inner flow in terms of a very simplified approach in
which we estimated the X-ray luminosity {\em within} our inner boundary (\S
\ref{sec:xray}). We found that in the past Sgr~A*'s X-ray luminosity could have been much higher than
its present value, and could have varied by several orders of
magnitude. Nethertheless, none of the peaks reached the $\sim 10^{39}$ \ergs
needed to explain the putative flare in Sgr A* X-ray luminosity a few hundred
years ago \citep[see also \cite{Muno07}]{Revnivtsev04}.  However, it is possible that a cold clump
falling into the inner region, such as those seen as peaks in Fig.~\ref{fig:acc_circ},
had a low enough angular momentum to circularise at a very small radius, $\sim 0.001"$. At that location, the clump mass would be higher than the
mass of the non-radiative flow by a few orders of magnitude.  Regardless of whether
the cold clump is evaporated by the hot flow, or sheared away and mixed with
it, in the end there would be much more mass in this region. The
accretion flow would then cool quickly, approaching 
a standard accretion disc configuration. The viscous time of this flow can be quite long,
of the order of 100 yr for a disc thickness $H/R \sim 0.01$,
making Sgr~A* stay in this state until the excess mass is accreted.  Assuming
a maximum radiative efficiency for this kind of flow, the bolometric
luminosity would be $\sim 10^{40}\,$\ergs, enough to give the observed
luminosity in the X-rays for the required period of time. We note that there were no
such low angular momentum cold clumps in the present simulations, but it is not
clear how robust this conclusion is with respect to changes in the model
parameters and resolution.

The wind properties we used are based on the spectroscopic study by
\cite{Martins07}. The uncertainty of their results is typically only of the
order of a few $\times$10\%, so we did not explore different realisations of the wind
data. There is more uncertainty in the wind properties from the 13E group, whose nature is
not clearly established yet, and 16SW, which is part of a binary
\citep{Martins06, Peeples07}. The analysis of these sources is not yet robust
enough and they are important contributors to the accretion. However, we
do not expect changes in the results to be larger than those resulting from
the different orbital configurations. We can still use our previous work
\citep{CNSD05, CNSD06, CN06} to understand the effect of changing the wind
properties. Clearly, larger mass-loss rates and slower wind velocities produce
more gas that is able to cool and form clumps. Too much of that gas can be
problematic in the sense that it would overproduce the atomic line emission
expected from gas at $T \sim 10^4\,$K. The same clumps increase the
variability of the accretion rate on time-scales of 10--100 yr.

Our results on the origin of the accreted material differ from those
obtained by \cite{Moscibrodzka06}. These authors argued that the accretion on
to Sgr~A* is dominated by the material expelled by 13E. One reason for the
discrepancy is that they used the older values for mass loss rates compiled by
\cite{Rockefeller04}, that give a rather extreme value for $\dot M_{\rm
13E}$. \cite{Moscibrodzka06} also neglected the orbital velocity of the
innermost stars, that are usually comparable to their wind velocity.

Other models in the literature have focused on the accretion of winds form the OB main sequence stars in the inner arc-second -- the `S-stars' \citep{Loeb04, Coker05}. While the inclusion of these stars would certainly make the models more realistic, their mass loss rates are probably at most $\sim 10^{-7}\msun\,$yr$^{-1}$. The star $\tau$~Sco has a spectral type (B0.2V) similar to S2 and the estimates of its mass loss rate range from $6 \times\ 10^{-8}\msun\,$yr$^{-1}$ \citep{Mokiem05} down to $< 6 \times\ 10^{-9} \msun\,$yr$^{-1}$ \citep{Zaal99}. Since mass loss rates are known to scale with luminosity \citep{Kudritzki00} and S2 is the brightest of the S stars, all of them should have very low $\dot{M}_{\rm w}$. Hence, we believe that the effect of the S stars would not affect our conclusions.

\section*{Acknowledgments} 
{
We acknowledge useful discussions with T.\ Paumard on the stellar orbits and the
observability of the gas emission, and with F.\ Yuan on RIAF models. 
We thank P.\ Ar\'evalo for proof-reading the draft and the referee for suggestions that significantly improved the paper.
JC acknowledges support from NASA
Beyond Einstein Foundation Science grant NNG05GI92G. FM acknowledges support from the Alexander von Humboldt foundation. The simulations
were performed on the JILA Keck-cluster, sponsored by the W.\ M.\ Keck
Foundation. Earlier calculations were ran at the Rechenzentrum Garching
and at the University of Leicester. JC acknowledges the hospitality of
the Theoretical Astrophysics group in Leicester, where part of this
work was done.
}

\bibliography{biblio.bib}
\bibliographystyle{mnras}

\label{lastpage}

\end{document}